\documentclass[reprint, superscriptaddress, amsmath, amssymb, prl,]{revtex4-1}

\usepackage{graphicx}
\usepackage{dcolumn}
\usepackage{bm}
\usepackage{color}

\begin{document}


\title{Direct Imaging of Charge Redistribution due to Bonding at Atomic Resolution via Electron Ptychography}
%
 \author{Gerardo T. Martinez}
 \thanks{These two authors contributed equally}
 \affiliation{%
 Department of Materials, University of Oxford, United Kingdom  
 }%
 \affiliation{%
 Materials and Component Analysis - Structural Analysis, Imec, Belgium  
 }%
 \author{Benjamin X. Shi}
 \thanks{These two authors contributed equally}
\affiliation{%
Department of Materials, University of Oxford, United Kingdom  
}%
\affiliation{Yusuf Hamied Department of Chemistry, University of Cambridge, United Kingdom}

 \author{Timothy C. Naginey}%
 \affiliation{%
 Department of Materials, University of Oxford, United Kingdom  
 }%

 \author{Lewys Jones}%
 \affiliation{
 School of Physics, CRANN and AMBER, Trinity College Dublin, the University of Dublin, Ireland}
 
 \author{Colum M. O'Leary}%
 \affiliation{%
 Department of Materials, University of Oxford, United Kingdom  
 }%
 \affiliation{Department of Physics and Astronomy, University of California, Los Angeles, USA}
 
 \author{Timothy J. Pennycook}%
 \affiliation{
 EMAT, University of Antwerp, Belgium}

 \author{Rebecca J. Nicholls}%
 \author{Jonathan R. Yates}%
 \author{Peter D. Nellist}%
 \email{peter.nellist@materials.ox.ac.uk}
\affiliation{%
 Department of Materials, University of Oxford, United Kingdom  
 }%


\date{\today}

\begin{abstract}
Phase imaging in electron microscopy is sensitive to the local potential, including charge redistribution from bonding. We demonstrate that electron ptychography provides the necessary sensitivity to detect this subtle effect by directly imaging the charge redistribution in single layer boron nitride. Residual aberrations can be measured and corrected post-collection, and simultaneous atomic number contrast imaging provides unambiguous sub-lattice identification. Density functional theory calculations confirm the detection of charge redistribution.
\end{abstract}

\maketitle


The electron charge density is of fundamental importance to the physics of materials. It is the redistribution of electrons that occurs when atoms bond that distinguishes materials from collections of independent atoms. The ground state electronic properties of any material system can be determined from the electron charge density, as stated by the Hohenberg--Kohn theorems \cite{PhysRev.136.B864,PhysRev.140.A1133}, the foundation on which density functional theory (DFT) is built. While DFT has been hugely successful as a tool for modeling materials systems quantum mechanically, the approximations used to make such calculations practical mean that there remains an important role for experiments to verify and inform DFT results. Therefore, there is great interest in experimental methods that can measure bonding and charge distributions locally. 

X-ray and electron diffraction are both sensitive to charge redistribution \cite{Koritsanszky2001,Wu1999,Shibata1999,Zuo1999}. Such diffraction experiments are, however, limited to periodic structures and do not provide local information, precluding the study of charge transfer around defects and interfaces. This is a major drawback as these features so often play a crucial role in the physics of materials systems. Furthermore, only the intensities of diffracted beams can be measured directly and the phase is lost, a loss of information known as the phase problem. For diffraction from crystals with an inversion centre the diffraction structure factor phases are trivial, being either zero or $\pi$ radians.  For non-centrosymmetric crystals, however, the structure factors are complex and determination of their phases is necessary to quantitatively measure bonding effects.  Mutual interference between multiply scattered beams in thicker samples, also known as dynamical scattering, does lead to diffracted intensities being dependent on structure factor phases, and this is exploited in the convergent beam electron diffraction (CBED) approach \cite{spence1993aca}. For thin samples, such as the monolayer used here, the required multiple scattering does not occur.  Scanning tunnelling microscopy provides spectroscopy that is sensitive to the local electronic environment, but it is limited to use with surfaces. Electron energy loss spectroscopy (EELS) is not limited to surfaces, can be performed at atomic resolution and can provide local electronic structure information via the interpretation of fine structure features  \cite{PhysRevB.79.085117,doi:10.1021/nn402489v}. However, such interpretation is an indirect method requiring matching of experimental spectra to forward simulations. 

If a lens is used to reinterfere the diffracted beams, as occurs in HRTEM, the phase of the diffracted beams is significant and controls the position of features in the image.  Phase contrast HRTEM is also one of the most dose efficient forms of local atomic resolution imaging, and its high phase sensitivity has been used to detect charge transfer at defects in graphene and in the polarized bonds of monolayer BN \cite{MEYER2011nmat}. However to form contrast, HRTEM requires either the intentional retention of lens aberrations, reducing resolution, or the use of a physical phase plate, which can suffer stability and charging issues \cite{DanevPNAS2014}. As a result, HRTEM imaging is highly sensitive to imaging parameters and is a potential cause for misinterpretation of the data. Phase-retrieval using off-axis holography is similarly limited. While accurate measurements of the mean inner potential can be made, detection of charge redistribution has proved elusive \cite{BorghardtPRL18}.  Thus a method is required that makes use of the phases of the diffracted beams to form a highly dose-efficient image that is sensitive to the electron charge density with robustness to residual aberrations.  
\begin{figure*}
    \centering
    \includegraphics{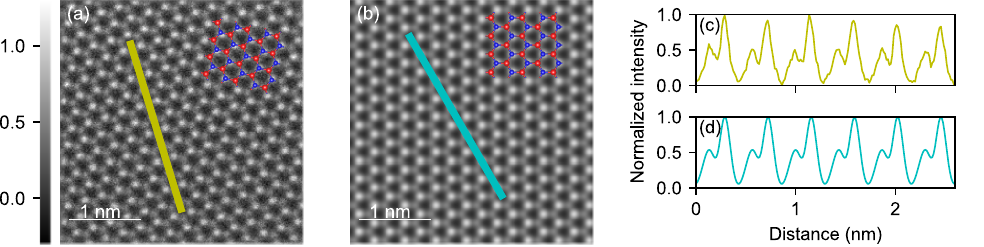}
    \caption{Annular dark field imaging of monolayer  hexagonal boron nitride using (a) template matching average of experimental data and (b) simulated data using the experimental conditions. Here, the intensities (given by the color bar) are normalized to the average peak pixel intensity for N. The line profiles are taken from the areas under the (c) yellow bar of the experimental image and the (d) blue bar of the simulated image. There was an average intensity ratio of 0.50 between B and N for experiment and 0.52 for simulation. N is blue and B is red in the overlaid model in (a) and (b). }
    \label{fig:figure1}
\end{figure*}

Here we demonstrate that phase imaging with electron ptychography \cite{HOPPE1969aAC,HOPPE1969bAC,HOPPE1969cAC,RODENBURG1992ptrslA,Rodenburg:1993tw} in the scanning transmission electron microscope (STEM) provides just such an optimized means of imaging local electronic charge densities. STEM ptychography is one of a number of emerging phase imaging methods in STEM, including differential phase contrast (DPC) \cite{ShibataNatPhys12}, integrated DPC \cite{LazicUltram16} and first-moment \cite{MuellerNatComm14} imaging.  Although originally developed as a means of achieving superresolution \cite{ NELLIST1995Nat, JiangNat2018}, the technique has recently been developed to provide high signal-to-noise phase images \cite{Pennycook:2015do,Yang:2015fg,YANG2016ncom, SongSciRep2019}. Most recently, electron ptychography has been shown to provide higher signal to noise images than phase contrast HRTEM for the same dose \cite{PENNYCOOK2019131}. No aberrations are required to form contrast in electron ptychography, the diffracted beam phase information is retrieved making use of the interference of overlapping diffracted discs, and once the phase problem is solved, the amplitude and phase of the scattering can be expressed in either reciprocal (diffraction) or real (imaging) space. Here we use a real-space representation to demonstrate a methodology that can also be applied to crystal defects.  Furthermore, by performing the experiments in focus in the STEM, one obtains simultaneously the high angle annular dark field (HAADF) signal. 
 
 The HAADF signal provides an independent image that is unaffected by the electron charge density, and contains strong compositional information. This provides the vital ability to accurately determine which elements are where in a material, and correctly interpret the phase images.  Moreover, although the ptychographic phase images are also sensitive to errors in the tuning of the aberrations, one can correct for such residual aberrations after taking the data \cite{YANG2016ncom}. As we show here, the ability to correct for such small errors in the electron optical aberration correction can prove crucial to achieving sufficient precision and accuracy to map the subtle effects of charge redistribution on the phase images.

For this study, a sample of monolayer hBN was imaged using a probe corrected JEOL ARM200CF microscope. The accelerating voltage was 80 kV with a 31.5 mrad semi-angle of convergence.  A set of 16 $512{\times}512$ probe position 4D data-sets was acquired using the JEOL 4DCanvas system incorporating a pnCCD device \cite{RYLLJInst2016} using $66{\times}264$ pixels in the detector plane at 4000 frames per second. 
The real space sampling in the image plane corresponds to a pixel size of $0.135~{\mbox \AA}$. 

Simultaneously acquired ADF image-series were used to diagnose environmental scanning-distortions using non-rigid registration (NRR) \cite{JONES2015asci}. These distortions were then compensated throughout each of the 16 4D data sets before their signals were summed. This 4D NRR both improves spatial precision (by a factor close to the square root of the number of input frames), and yields a single 4D volume with a greater signal to noise ratio; this in-turn leads to a more precise aberration diagnosis and correction, as well as an increase in the attainable phase-precision compared to that which could be obtained from just a single scan.  A high signal to noise ratio ADF image is also output [Fig.~\ref{fig:figure1}(a)], which is used for the unambiguous identification of the two sub-lattices.  The average dose-per-frame was calculated to be $\mathrm{7.75 {\times} 10^5~ e^-/{\mbox \AA}^{2}}$, which leads to a total dose for the non-rigid aligned 4D data-set of $\mathrm{1.24 {\times} 10^7~e^-/ {\mbox \AA}^{2}}$. The ADF image also provides confirmation that the sample is indeed a monolayer. The ratio of the B and N site intensities in the ADF image matches that of the single layer simulation, as shown in Fig.~\ref{fig:figure1}, but not those of the thicker multilayer structures in which the difference between the two sites decreases significantly (see the supplemental information for further details).

\begin{figure*}[t!]
    \centering
    \includegraphics[width=0.9\textwidth]{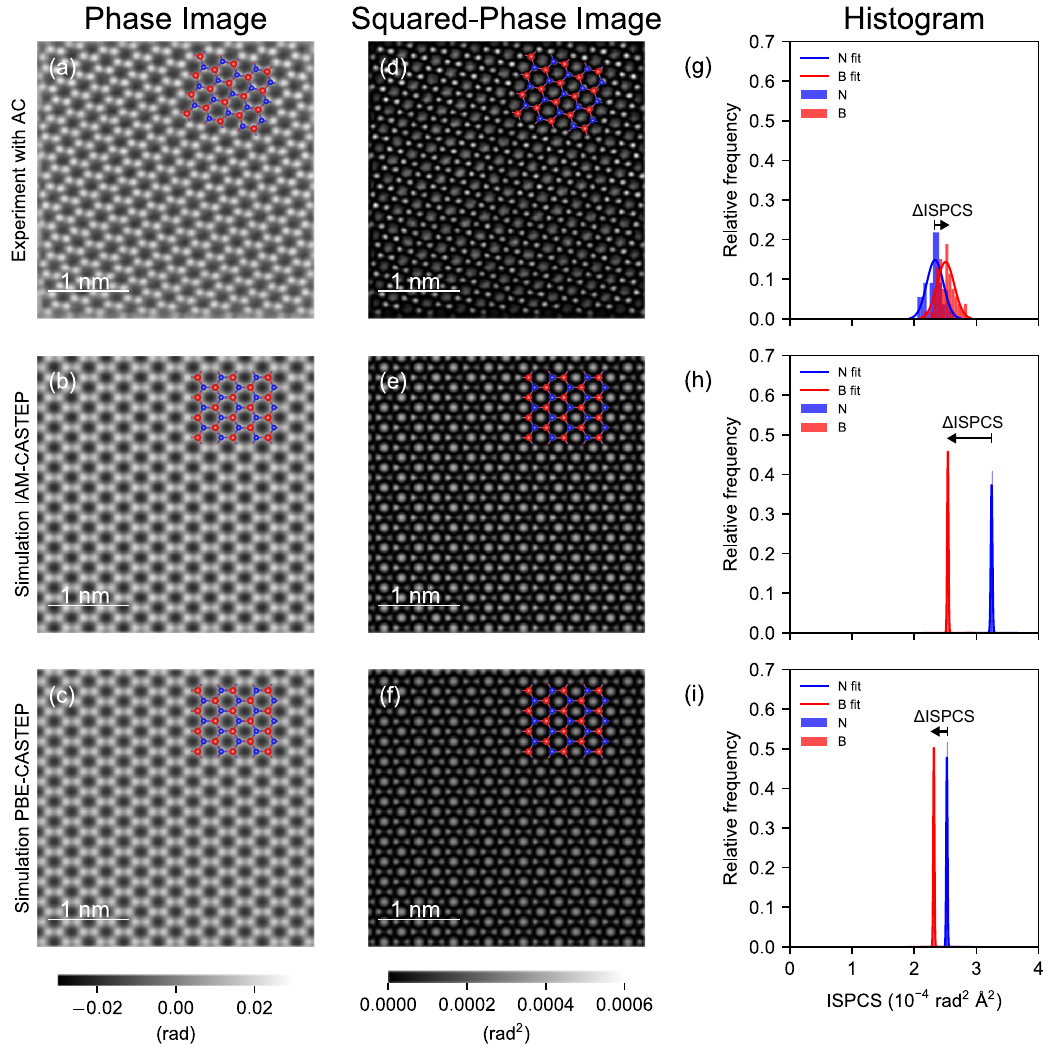}
    \caption{The WDD phase image of hBN from experimental data (a) with post-collection aberration correction (AC) is compared to images from simulated data using (b) IAM and (c) DFT-CASTEP using the PBE functional. (d) to (f) show the corresponding squared-phase images of the images in (a) to (c), and (g) to (i) show the corresponding histograms of the ISPCS of N and B with respect to vacuum obtained from the parametric model fitted to the squared-phase images.}
    \label{fig:figure2}
\end{figure*}

The non-rigid aligned ptychographic data-set was then reconstructed using the Wigner distribution deconvolution (WDD) method \cite{RODENBURG1992ptrslA} and the probe aberrations were determined using the method described in \cite{YANG2016ncom}. Because an inline technique such as ptychography only measures relative phase changes, the retrieved phase images were plotted with the mean phase across the entire image set to zero. Similarly, the lack of low-frequency transfer results in a point-spread function that contains positive and negative regions. To give a metric allowing comparison with simulation, the resulting phase images were then squared to give the squared-phase image and analyzed using a modified version of the Absolute Integrator Software \cite{JONESAI}. Voronoi cells at each atomic sites provided an area over which the squared-phase signal was integrated. We will refer to this quantity as the integrated squared-phase cross-section (ISPCS). The ISPCS was found to be more robust to residual aberrations than the peak phase value (see supplemental information). The distribution of ISPCS values for each of the B and N sites can be examined through statistical analysis. For example, Fig.~\ref{fig:figure2}(g) shows the histogram of the ISPCS for the two atom types (N - blue and B - red) for the ptychographic phase image reconstruction after  correction of residual aberrations. The mean ISPCS values for each atom type are quoted in Table~\ref{tab1:hbnispcsresults} along with their precisions, which have been taken to be the sum of the standard error of the mean together with an error arising from uncertainties in atomic positions. The distributions for each atom type are also fitted with Gaussian functions as an aid to the eye. All the images presented in Fig.~\ref{fig:figure2} were analysed in the same manner.

In hBN, the transfer of electrons occurs from B to N sites, leading to an increased screening of the potentials of the N nuclei \cite{PAULING1966pnas}. This effect induces a smaller phase-shift in the electron wave than the neutral atom potential.  Table~\ref{tab1:hbnispcsresults} shows that $\Delta$ISPCS, the difference between the mean ISPCS of N and B, is $-0.218 \times 10^{-4}$~rad\textsuperscript{2}~\AA\textsuperscript{2} for the reconstructed phase image without ptychographic aberration correction, which is refined to $-0.168 \times 10^{-4}$~rad\textsuperscript{2}~\AA\textsuperscript{2} after performing ptychographic reconstruction [Fig.~\ref{fig:figure2}(a), (d) and (g)]. It is therefore clear that the small residual aberrations, although not observable in the ADF image, are still significant at the precision of measurement offered by ptychography. In Fig.~\ref{fig:figure2} we compare the experimental results with image simulations. We used an in-house code to simulate the CBED patterns that were used to obtain ptychographic phase images. An electron source probe broadening of FWHM 1.02 \AA\ was used, as this was determined to match the mean ISPCS of all N and B atoms in the simulated phase image (using the PBE exchange-correlation functional \cite{pbe}) to the experimental phase image. The same source size effect was included in the ADF simulations which also match the experimental data, offering confidence in this value. Figures \ref{fig:figure2}(b), (e) and (h) show the results when simulating a 4D dataset using the independent atom model (IAM) that does not include bonding. Our IAM potentials are obtained from DFT calculations with isolated atoms using the PBE functional. By using DFT to obtain the IAM potentials, we avoid the need to use parametrized IAM potentials which are commonly used in electron microscopy simulation software \cite{LOBATO2014aca}. As can be observed in Fig.~\ref{fig:figure2}(h), the mean difference between the ISPCS values for the two species is $0.704 \times 10^{-4}$~rad\textsuperscript{2}~\AA\textsuperscript{2} using the IAM potentials, far greater than the experimental result. It can be inferred that the experimental results indicate the direct observation of charge transfer and the necessity of simulations that include bonding effects. 

\begin{table}[h!]
	\caption{Calculated hBN $\Delta$ISPCS for IAM, experiment and several density functional approximations. IAM and DFT simulations used CASTEP unless stated otherwise.}
	\label{tab1:hbnispcsresults}
	\begin{ruledtabular}
	\begin{tabular}{l c c}
		&  \multicolumn{1}{c}{\textrm{$\Delta$ISPCS}}                                                                 & \multicolumn{1}{c}{Mulliken charge transfer}  \\ 
		& \multicolumn{1}{c}{\textrm{($10^{-4}$ rad\textsuperscript{2} \AA\textsuperscript{2})}}  & \multicolumn{1}{c}{(electrons)} \\ \hline
		IAM          &  \hphantom{$-$}$0.704 \pm 0.007$ &                          \\
		LDA            &  \hphantom{$-$}$0.234\pm 0.007$ & 0.84                     \\
		PBE            &  \hphantom{$-$}$0.210\pm 0.007$ & 0.86                     \\
		PBE (WIEN2k)       &  \hphantom{$-$}$0.204\pm 0.007$ &                          \\
		rSCAN          &  \hphantom{$-$}$0.137\pm 0.007$ & 0.92                     \\
		PBE0          &  \hphantom{$-$}$0.111\pm 0.007$ & 0.95                     \\
		Expt.\ (with AC)            &  $-0.168\pm 0.043$ &                                            \\ 
		Expt.\ (no AC) &  $-0.218 \pm 0.032$    &   \\
	\end{tabular}
	\end{ruledtabular}
\end{table}

To include the effects of bonding, we used projected potentials obtained from two different materials modeling codes; the plane-wave pseudopotential code CASTEP \cite{CASTEP999} and the all-electron LAPW+lo code WIEN2k \cite{wien2k}. A modified version of the wien2venus script \cite{wien2venus} was used to calculate projected potentials from WIEN2k. To compute the projected potentials from CASTEP, it is necessary to correctly account for the use of pseudopotentials and the details of our implementation are reported elsewhere \cite{TCN_METHOD}. Calculations were performed on an orthorhombic unit cell of monolayer hBN containing 4 atoms and with lattice parameters of $a=2.50~{\mbox \AA}, b=4.34~{\mbox \AA},$ and $c=21.17~{\mbox \AA}$. The WIEN2k SCF calculation was performed using a value of 7 for the quantity $R_{MT} \cdot K_{max}$. For the SCF calculation in CASTEP, a planewave cutoff energy of 1100 eV was used, along with pseudopotentials generated using the ultrasoft scheme \cite{vanderbilt-prb90}. Both calculations used a maximum k-point spacing of $0.095~{\mbox \AA}^{-1}$ and the PBE functional \cite{pbe}, with additional calculations also performed using the LDA, rSCAN and PBE0 functionals in CASTEP. All image simulations were performed using the same source broadening of 1.02~\AA. The resulting phase image for PBE-CASTEP is shown in Fig.~\ref{fig:figure2}(c) with its corresponding ISPCS histograms in (i). The analysis of these simulations (Table~\ref{tab1:hbnispcsresults}) shows that the results between WIEN2k and CASTEP simulations for PBE are within the statistical error of each other, consistent with the findings of a recent comparison \cite{SUSI201916} of results from WIEN2k and the projector-augmented wave method code GPAW. The $\Delta$ISPCS value for the PBE-CASTEP calculation is $0.210 \times 10^{-4}$~rad\textsuperscript{2}~\AA\textsuperscript{2}, much closer to the $-0.218 \times 10^{-4}$~rad\textsuperscript{2}~\AA\textsuperscript{2} observed experimentally than the $0.704 \times 10^{-4}$~rad\textsuperscript{2}~\AA\textsuperscript{2} value for the IAM. The use of functionals with more accurate electron densities (PBE0 and rSCAN) can bring the $\Delta$ISPCS value even closer to experiment, whilst the less accurate functionals (LDA) brings $\Delta$ISPCS further from experiment with respect to PBE. Mulliken population analysis suggests that a decrease in $\Delta$ISPCS corresponds to an increase in charge transfer from B to N. The DFT calculations can therefore explain about 65\% of the change in the ISPCS values seen experimentally, but there remains a statistically significant mismatch.

In conclusion, we have demonstrated the experimental observation of charge transfer in hexagonal boron nitride using non-rigid aligned aberration-free phase imaging by means of WDD electron ptychography. The use of ptychographic post-collection aberration correction ensures the phase measurements accurately reflect the influence of bonding. Only simulations that included the effects of bonding via DFT match the experimental images. Furthermore, by performing the phase imaging in focus in STEM, we have shown how the simultaneous ADF signal allows one to unambiguously identify the different elements, averting misinterpretation of the phase images. 


We acknowledge funding from the EPSRC (grant numbers EP/M010708/1,
EP/K032518/1, EP/K040375/1 and EP/L022907/1). The authors are thankful to Y.~Sasaki from JFCC for providing the sample for this study. Technical support and fruitful discussions with Y.~Kondo and R.~Sagawa from JEOL Ltd and M.~Huth, M.~Simsom and R.~Ritz from PNDetector GmbH and H.~Soltau from PNSensor GmbH in the implementation of 4D Canvas system are also greatly appreciated. TCN is grateful for support from the EPSRC Centre for Doctoral training, Theory and Modelling in Chemical Sciences, under grant EP/L015722/1.  LJ is supported by SFI (grant numbers URF/RI/191637 and 19/FFP/6813). T.J.P acknowledges funding from the European Union's Horizon 2020 Research and Innovation Programme under the Marie Sk\l odowska-Curie grant agreement no.~655760--DIGIPHASE and European Research Council Grant No.~802123-HDEM.

\end{document}



\title{Supplemental information: Direct imaging of charge redistribution in monolayer hexagonal boron nitride using electron ptychography in the scanning transmission electron microscope}

 \author{Gerardo T. Martinez}
 \thanks{These two authors contributed equally}
\affiliation{%
 Department of Materials, University of Oxford, United Kingdom  
 }%
 \affiliation{%
 Materials and Component Analysis - Structural Analysis, Imec, Belgium  
 }%
 
\author{Benjamin X. Shi}
\thanks{These two authors contributed equally}
\affiliation{%
Department of Materials, University of Oxford, United Kingdom  
}%
\affiliation{Yusuf Hamied Department of Chemistry, University of Cambridge, Lensfield Road, Cambridge CB2 1EW, United Kingdom}
 \author{T C Naginey}%
 \affiliation{%
 Department of Materials, University of Oxford, United Kingdom  
 }%
 
 \author{L Jones}%
 \affiliation{
 School of Physics and CRANN,Trinity College Dublin, the University of Dublin, Ireland}
 
 \author{C O'Leary}%
\affiliation{%
 Department of Materials, University of Oxford, United Kingdom  
 }%
 \affiliation{Department of Physics and Astronomy, University of California, Los Angeles, USA}
 
 \author{T J Pennycook}%
 \affiliation{
 EMAT, University of Antwerp, Belgium}
 
 \author{R J Nicholls}%
 \author{J R Yates}%
 \author{P D Nellist}%
 \email{peter.nellist@materials.ox.ac.uk}
\affiliation{%
 Department of Materials, University of Oxford, United Kingdom  
 }%

\date{\today}


\maketitle
\renewcommand{\thefigure}{S\arabic{figure}}

\begin{figure}
    \centering
    \includegraphics{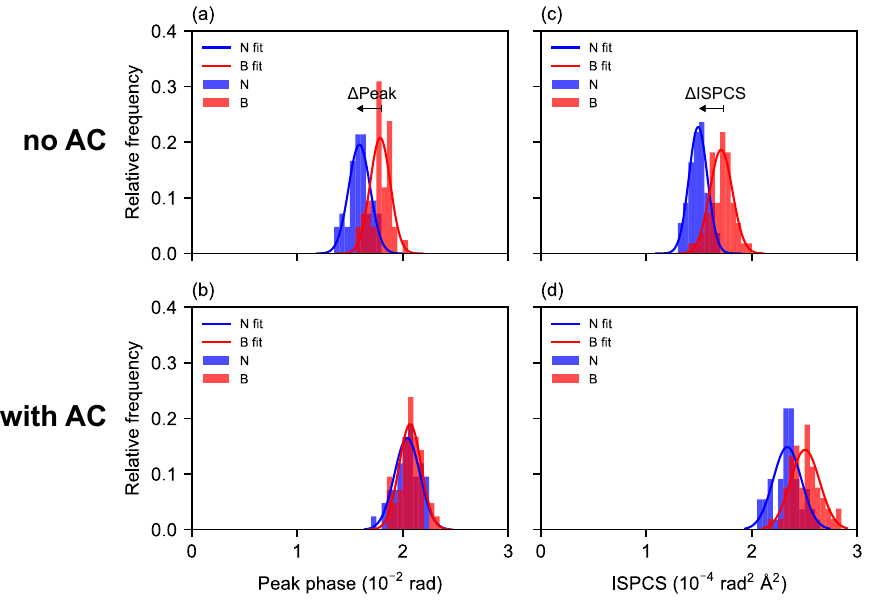}
    \caption{Comparison of ISPCS vs peak phase $\Delta$ difference (relative to the vacuum) values for non-aberration corrected and aberration corrected WDD electron ptychography. Histograms show the values of N (blue) and B (red) sites. 
    a) and b) correspond to non-aberration corrected case, while c) and d) show the results of aberration correction up to 2nd order. }
    \label{fig:supp_figureA}
\end{figure}

\begin{samepage}

\section{Robustness of ISPCS vs Phase peak values}

Integrated squared-phase cross-sections (ISPCSs) were used rather than peak phases to provide a metric to compare with simulated images that is more robust to any uncorrected, residual aberrations.  The robustness can be seen in Fig.~\ref{fig:supp_figureA} where the correction of aberrations (values shown in Table \ref{table:suptab1}) during the ptychography reconstruction causes smaller changes in the ISPCSs of the B and N atoms than occurs for the peak phases.  The enhanced robustness is because part of the effect of residual aberrations will be to broaden the peaks in the phase image by redistributing the pixel values.  Similar to the discussion in \cite{E2013Ultram} for annular dark-field imaging, the integration over a peak to form a cross-section captures the redistributed values.

\end{samepage}

\begin{table}[h!] 
\caption{\label{table:suptab1} Measured residual aberrations in the experimental phase images.}
\begin{ruledtabular}
\begin{tabular}{ccccccc}
C1 (nm)      & C12a (nm)      & C12b (nm)      & C23a (nm)      & C23b (nm)      & C21a (nm)       & C21b (nm)     \\ \hline
0.384 & -0.331 & -0.409 & 17.012 & -14.815 & -70.026 & 1.039
\end{tabular}
\end{ruledtabular}
\end{table}

\section{Confirmation of hBN monolayer}

\begin{figure}[h]
    \centering
    \includegraphics{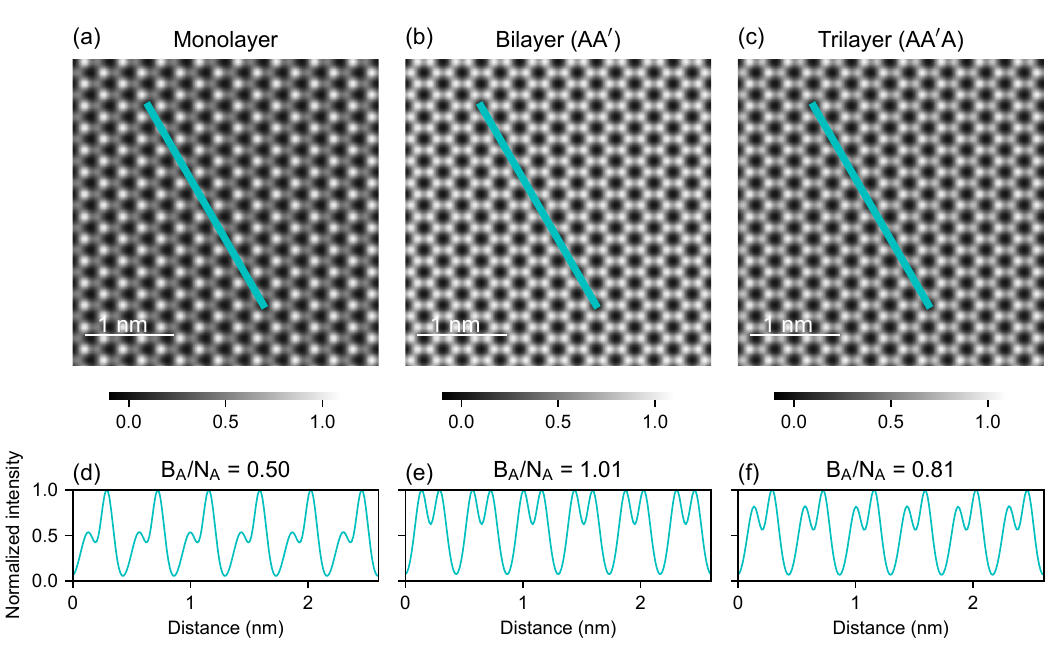}
    \caption{HAADF STEM image simulations of a hBN (a) monolayer, (b) bilayer and (c) trilayer with their corresponding profiles in (d), (e) and (f) respectively. The profiles were generated by integrating a 7 pixels wide ($0.95 {\mbox \AA}$) line shown in cyan.}
    \label{fig:supp_figureB}
\end{figure}

ADF STEM image simulations using the experimental conditions were performed in order to confirm that the experimental data was indeed a monolayer of hBN. The simulations were carried out in CASTEP, in a similar manner to the ptychograpy simulations in the main text, by using the weak phase object approximation (WPOA)~\cite{kirklandAdvancedComputingElectron2010b} to produce CBED patterns from the electrostatic potential. The simulated images were normalised in order to perform a relative comparison of the scattered intensities of N and B atoms. The normalisation consisted of subtracting a constant background so that the hexagonal holes in between atoms would have a mean value of 0 and then dividing by the mean value of the N atomic sites. The mean peak intensity of the N sites is then 1. Figures \ref{fig:supp_figureB} (a), (b) and (c) show the simulated images of a monolayer, bilayer and trilayer of hBN respectively. Their corresponding line profiles are shown in Fig.~\ref{fig:supp_figureB}(d), (e) and (f). These profiles integrate the scattered intensity over the width of the atoms, which for this case is 7 pixels ($0.95~ {\mbox \AA}$). The ratios of the atomic columns containing the B and N atoms in the A layer are 0.50, 1.01 and 0.81 for a monolayer, AA$^{\prime}$ bilayer and AA$^{\prime}$A trilayer respectively.  The experimental value is 0.52, confirming monolayer hBN.

\section{Phase image comparison using IAM and DFT approaches}

\begin{figure}
   \centering
    \includegraphics[width=\textwidth]{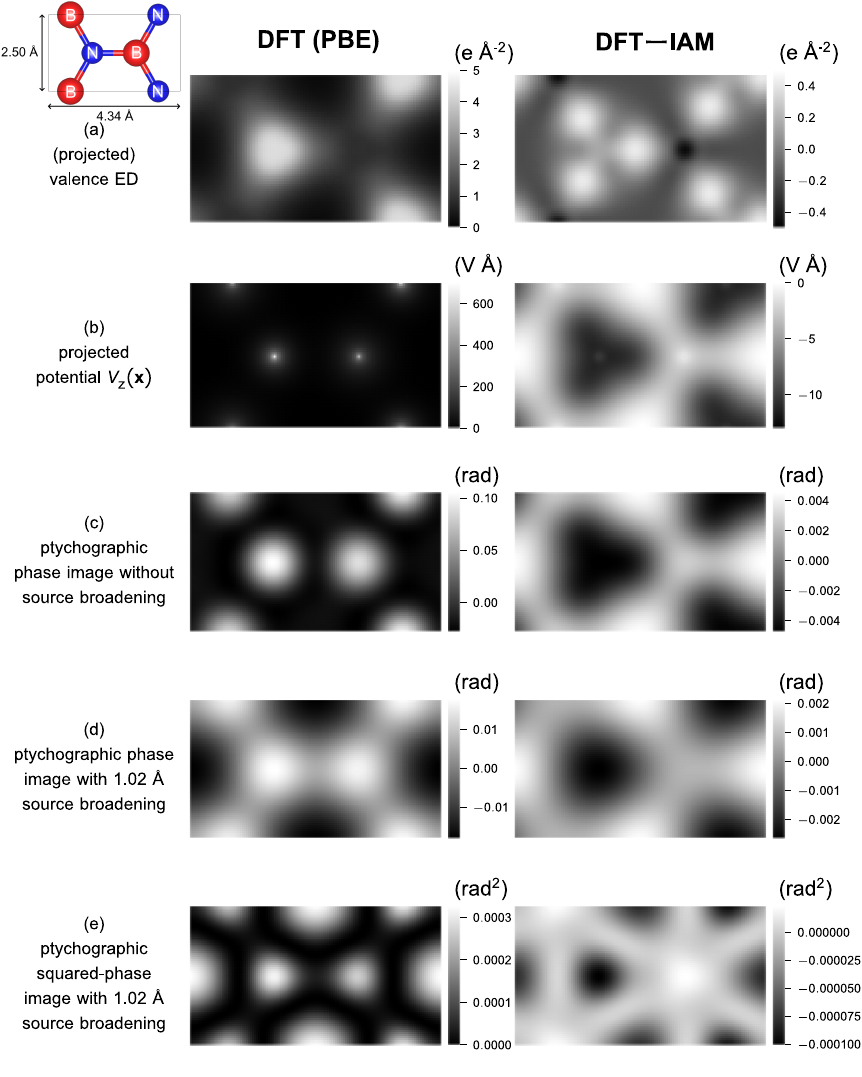}
    \caption{PBE-DFT (left) and its difference with the IAM (right) in hBN for several quantities in the ptychographic phase reconstruction workflow.}
    \label{fig:supp_figureC}
\end{figure}

Figure \ref{fig:supp_figureC} shows the difference between the phase images obtained from simulating the hBN potential using the IAM and DFT approaches. These images correspond to Fig. 2 (b) and (c). An overlay of the atoms positions have been depicted in the figure, in which N atoms correspond to black dots and B atoms correspond to white dots. As observed in Fig.~\ref{fig:supp_figureC}, the main phase changes occur close to the atom sites, particularly at the N atom positions. The localisation of the phase changes close to the atom sites also supports the procedure of integrating the phase over the atom sites, that is, the ISPCS. By doing so, we can account for the charge transfer in a robust way. 
\clearpage

\clearpage




\bibliography{bibliography.bib}